\documentclass[prl,aps,twocolumn,showpacs,superscriptaddress]{revtex4}
\usepackage{graphicx}
\usepackage{amsmath,amssymb,revsymb}
\usepackage[svgnames]{xcolor}
\usepackage[colorlinks=true,citecolor=Purple,filecolor=Purple,urlcolor=Purple,linkcolor=Purple]{hyperref}
\begin{document}

\title{Efimov physics in heteronuclear four-body systems}
\author{Yujun Wang}
\altaffiliation{Present address: JILA, University of Colorado, 440 UCB, Boulder, Colorado, 80309, USA}
\affiliation{Department of Physics, Kansas State University, Manhattan, Kansas 66506, USA}
\author{W. Blake Laing}
\altaffiliation{Present address: Department of Physics and Astronomy, Rowan University, Glassboro, New Jersey 08028, USA}
\affiliation{Department of Physics, Kansas State University, Manhattan, Kansas 66506, USA}
\author{Javier von Stecher}
\affiliation{JILA, University of Colorado, 440 UCB, Boulder, Colorado, 80309, USA}
\author{B.D. Esry}
\affiliation{Department of Physics, Kansas State University, Manhattan, Kansas 66506, USA}

\begin{abstract}
We study three- and four-body Efimov physics in a heteronuclear atomic system with three 
identical heavy bosonic atoms and one light atom. 
We show that exchange of the light atom between the 
heavy atoms leads to both three- and four-body features in
the low-energy inelastic rate constants that trace to the Efimov effect.  Further, the effective interaction
generated by this exchange can provide an additional mechanism for control in ultracold
experiments. Finally, we find that there is no true four-body Efimov effect
--- that is, no infinite number of four-body states in the absence of two- and three-body
bound states --- resolving a decades-long controversy.
\end{abstract}

\maketitle

Few-body physics has benefitted greatly from ultracold experiments on quantum gases in recent years.
Many long-standing predictions have been verified and new phenomena identified due to 
the extraordinary ability 
to control and measure these systems.  One of the more spectacular results
for few-body physics was the observation of Efimov physics in Cs~\cite{Kraemer2006}, experimentally validating
Efimov's surprising, 
35-year old prediction~\cite{Efimov1970}
by confirming its predicted influence on 
three-body recombination~\cite{Esry1999}.  Several measurements of Efimov physics have now been 
made~\cite{Kraemer2006,Knoop2009,Ottenstein2008,Huckans2009,Barontini2009,Zaccanti2009,Gross2009,Williams2009,Pollack2009,Lompe2010,Berninger2011}
and even include features traceable via theory~\cite{4BStecher2009} to four-body processes~\cite{4BFerlaino2009,Pollack2009}.

One natural question to ask is whether there is an Efimov effect for $N$$>$3 bodies.  
The Efimov effect in this case is defined as the existence of an infinite number of $N$-body
bound states when no subsystems are bound. 
Part of the answer was provided
in Ref.~\cite{Amado1973}:
there is no Efimov effect for $N$$>$3 
{\em equal mass} particles. 
While generally interpreted to imply no Efimov effect is possible 
for $N$$>$3 
(see, for example, \cite{Braaten2006}), this result does not preclude 
the possibility of an Efimov effect for systems with 
\emph{unequal masses}.
In fact, Ref.~\cite{CastinFourBody} recently reported that three identical
fermions interacting resonantly with a fourth particle do have an Efimov effect
for a small range of mass ratios.
The question of an Efimov effect for $N$=4 with three identical bosons, however, remains open
despite prior study:
$H_3L$ systems ($H$ and $L$ are heavy and light particles, respectively)
were analyzed in Ref.~\cite{Fonseca1981} with the conclusion
that no Efimov effect occurs.  When this system was revisited in Ref.~\cite{Tjon1987}, however,
the opposite conclusion was reached.

In this Letter, we settle this controversy:
there is no true four-body Efimov 
effect for $H_3L$ with bosonic $H$s.  Although our conclusion is
in agreement with Ref.~\cite{Fonseca1981}, our reasoning is very different.  We have, however, identified one universal four-body state 
attached to each three-body Efimov threshold. We also show that 
the low-energy scattering observables simultaneously display distinct three- and four-body features characteristic of
the three-body Efimov effect.  Moreover, we find that, in the context of ultracold
collisions, the $s$-wave two-body scattering length $a_{H\!L}$ between $H$ and $L$ 
atoms can be used to tune the effective heavy-heavy scattering length $a_{H\!H}^*$ when $a_{H\!L}>0$, opening up 
new avenues for control in few-body systems.  

Since a direct solution of the four-body problem remains a substantial challenge,
we apply the Born-Oppenheimer (BO) approximation as described in 
Refs.~\cite{Fonseca1981,Tjon1987}, 
assuming that the mass $m$ of $L$ is much less than the mass $M$ of $H$.  
Although we will focus on bosonic $H$s, our analysis can be generalized straightforwardly.

We first 
recall that the three-body Efimov effect occurs when at least two 
scattering lengths are much larger than a characteristic two-body interaction range $r_0$~\cite{Efimov1970}. 
Thus, for $H_2L$, 
Efimov physics
will occur for $|a_{H\!L}|$$\gg$$r_0$.
Applying the BO approximation with $\rho$ the $H$+$H$ distance yields a BO potential with
the universal long-range behavior $-\chi_0^2\hbar^2/2m\rho^2$, $\chi_0^{~}$$\approx$0.567143, in the region 
$r_0$$\ll$$ \rho$$\ll$$ |a_{H\!L}|$~\cite{Fonseca1979,Braaten2006}.  The three-body problem has thus been reduced to
an effective two-body problem whose interaction is 
the Efimov potential, giving the characteristic geometric relation between bound state energies
\begin{eqnarray}
E_{n+1}/E_n=e^{-2\pi/s_0},  \qquad n=0,1,2,...
\label{Eq_EfimovScale}
\end{eqnarray}
where $s_0^2$=$\chi_0^2M/2m-1/4$.
An infinite number of Efimov states are thus obtained
in the limit $|a_{H\!L}|$$\rightarrow$$\infty$, corresponding to a zero-energy $H\!L$ bound state.

Similarly, all of the universal results~\cite{Braaten2006} for low-energy three-body scattering apply to $H_2L$.
For instance, when $a_{H\!L}$$>$0, there is a weakly bound $H\!L$ molecule, and 
\begin{eqnarray}
a_{H\!H}^*\propto \cot[s_0\ln(a_{H\!L}/r_0)+\Phi]a_{H\!L}
\label{Eq_aHH}
\end{eqnarray}
where $r_0$ is the characteristic range of the two-body interaction.
The poles in $a_{H\!H}^*$ occur at those values of $a_{H\!L}$ when an Efimov state becomes
bound, so their positions are characteristically log-periodic in $a_{H\!L}$.
Their overall position, however, is determined by
the short-range three-body phase $\Phi$~\cite{Esry1999,PhaseDIncao}.  Equation~(\ref{Eq_aHH})
thus shows that $a_{H\!L}$ 
provides a control over $a_{H\!H}^*$ that could 
prove advantageous when $a_{H\!H}^*$ cannot easily
be controlled directly via, for example, 
a Feshbach resonance~\cite{ChinJulienneRMP}. 

In general, the $H$s also interact directly. However, so long as the direct $H$+$H$ interaction is short-ranged --- 
even if it is repulsive --- the large-$\rho$ behavior of the effective $H$+$H$ interaction is unchanged,
and the Efimov effect remains.  A direct $H$+$H$ interaction
changes only $E_0$ (or, equivalently, $\Phi$) and not the characteristic log-periodic behavior.  
Consequently, manipulating the direct $H$+$H$ interaction allows control of $\Phi$ 
and thus the position of the family of log-periodic Efimov features.  

Applying the BO approximation to $H_3L$ reduces it to an effective three-body problem for $H_3$.
We thus expect that our knowledge of Efimov physics
and universality for three identical bosons should apply and
provide at least a basic understanding of the system.
For simplicity, we 
assume that the $H$ atoms do not interact directly.  Per the argument above,
we know this will not materially affect our conclusions. 
The $H$+$H$ interaction then comes solely from mediation by the $L$ and is
characterized by the effective $H$+$H$ scattering length $a_{H\!H}^*$ from Eq.~(\ref{Eq_aHH}).
Based on the result for three bosons, we expect an Efimov effect for $H_3$ when $|a_{H\!H}^*|$$\rightarrow$$\infty$~\cite{Tjon1987}.

To determine whether the $|a_{H\!H}^*|$$\rightarrow$$\infty$ limit 
actually 
produces an Efimov effect for $H_3L$,
we adopt 
the following definition~\cite{Amado1973,Fonseca1981}: 
a true 
four-body Efimov effect exists if the four-body system possesses an infinity 
of stable bound states when there is a 
zero-energy three-body bound state and no other two- or three-body bound states.
Under these conditions, the four-body Efimov effect is completely independent of 
any three-body Efimov effect, making the resulting states fundamentally different from the universal four-body states 
discussed in Refs.~\cite{4BPlatter2004,4BHammer2007,4BStecher2009,Schmidt2010}.

If it exists, a true four-body $H_3L$ Efimov effect will occur for
$|a_{H\!H}^*|$$\rightarrow$$\infty$ and $a_{H\!L}$$<$0.
This case gives a zero-energy $H_2L$ bound state but no $H\!L$ bound state and
is precisely the case identified in Ref.~\cite{Fonseca1981}.
They concluded that no Efimov effect is possible because the BO potential surface 
has no long-range component.  
Naus and Tjon~\cite{Tjon1987} correctly pointed out that it is not the BO
potential surface that must behave as $-R^{-2}$, but rather the adiabatic hyperspherical potential
where the hyperradius $R$ measures the overall size of the system
(see, for example, Refs.~\cite{NielsenReview,HeCalc2} for a discussion of hyperspherical coordinates in this context). 
They concluded that because $|a_{H\!H}^*|$$\rightarrow$$\infty$, the conditions of the Efimov effect
for $H_3$
are fulfilled and there is thus a four-body Efimov effect.
Unfortunately, neither analysis provides a definitive answer. 

Although otherwise sound, the argument of Ref.~\cite{Tjon1987} is limited by its
reliance on the BO approximation --- a problem they identified but did not address.
The issue with the BO approximation is that for $a_{H\!L}$$<$0, $L$ becomes unbound 
when $\rho$$\gtrsim$$|a_{H\!L}|$~\cite{Fonseca1981},
and its spectrum becomes continuous.  
Since a zero-energy $L$ is no longer fast compared to the $H$s,
the BO approximation breaks down.  Naus and Tjon 
followed Ref.~\cite{Fonseca1981} and simply set the BO potential at such points to zero. 
Under this questionable approximation, $a_{H\!H}^*$
can still be defined and adjusted to infinity, so that they conclude there is
an infinite series of bound $H_3L$ states, i.e. the four-body Efimov effect.

Where the BO approximation fails, however, the adiabatic hyperspherical approximation
is perfectly valid. 
In particular, the lowest $H_2L$ 
adiabatic hyperspherical potential, 
which corresponds to the BO potential, crosses
zero energy at a hyperradius $R$$\sim$$|a_{H\!L}|$ much like the BO potential, but 
remains discrete and 
increases to a barrier with height $\sim$1/$a_{H\!L}^2$~\cite{EsryJPBLetter}. 
For $R$ beyond this barrier, the
potential falls to zero as $l_{\rm eff} (l_{\rm eff}\!+\!1)\hbar^2/M R^{2}$ with $l_{\rm eff}$=3/2~\cite{Esry2001}. 

Having taken care to define the $H_2L$ hyperradius $R$ such that it reduces to the $H$+$H$ distance $\rho$
in the $M/m$$\rightarrow$$\infty$ limit, we can regard this adiabatic hyperspherical potential as an effective $H$+$H$ interaction.
Since the Efimov effect for the three $H$s can only occur for $s$-wave two-body interactions,
{\em i.e.} $l_{\rm eff}$=0, the fact that $l_{\rm eff}$=3/2 prohibits the Efimov effect.
This conclusion has been confirmed by direct calculations of the four-body adiabatic hyperspherical
potentials for $H_3L$ with $M/m$=30~\cite{StecherUnpub} using the correlated Gaussian approach~\cite{StecherCG}.

Even though our finding no true four-body Efimov effect agrees with the conclusions of 
Ref.~\cite{Fonseca1981} and contradicts the conclusions of Ref.~\cite{Tjon1987}, we believe
the former were right for the wrong reason and the latter underestimated the consequences
of the breakdown of the BO approximation.  In the end, it is this breakdown that 
excludes the possibility of a true four-body Efimov effect in this system.

\begin{figure}
\includegraphics[clip=true,width=0.95\columnwidth]{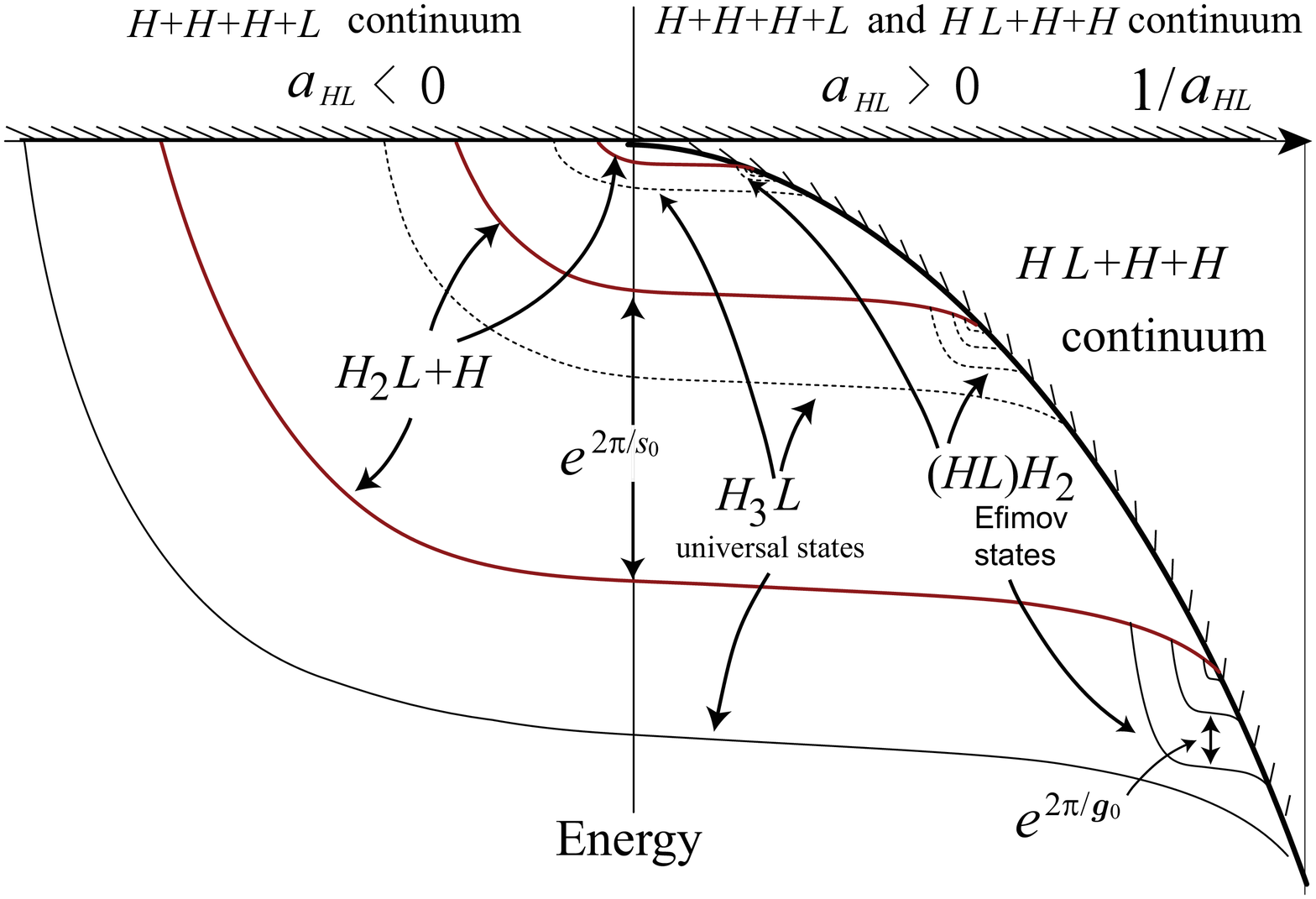}
\caption
{
Schematic energy spectrum for $H_3L$. 
Solid lines denote bound states and dashed lines denote resonances.
}
\label{Fig:4BSpec}
\end{figure}

The possibility of true four-body Efimov states is not the only phenomenon of interest
in $H_3L$.  
Continuing to $a_{H\!L}$$>$0 such that $a_{H\!L}$$\gg$$r_0$,
the BO approximation displays no pathologies since $L$ 
is bound for all $H$ configurations, and we can safely think about the three-body $H_3$
motion on the lowest BO potential surface.  
Based on the known three-body results~\cite{Efimov1973,NielsenReview,Braaten2006}, 
we expect an adiabatic hyperspherical potential
of the form
\begin{eqnarray}
W_0=-\frac{g_0^2+1/4}{2\mu R^2}\hbar^2, \quad a_{H\!L}\ll R\ll |a_{H\!H}^*|,
\label{Eq_EfimovPot}
\end{eqnarray}
with $g_0$=1.00624 
and $\mu$=$M$/$\sqrt{3}$ 
that approaches the three-body break-up threshold.  In this
case, that threshold corresponds physically to $H\!L$+$H$+$H$.
The lower limit of $R$ in Eq.~(\ref{Eq_EfimovPot}) is modified from 
the usual three-body problem~\cite{Efimov1973} due to the fact that the characteristic
range of the effective two-body $H$+$H$ potential is no longer $r_0$, but rather $a_{H\!L}$
as defined by the size of the $H\!L$ bound state.

When $a_{H\!L}$ is tuned to give $a_{H\!H}^*$$\rightarrow$$\infty$, the Efimov
potential (\ref{Eq_EfimovPot}) extends to infinity,
producing an infinite series of four-body bound states below the $H\!L$+$H$+$H$ threshold with 
binding energies $E_{n+1}/E_n$=$e^{-2\pi/g_0}$.
These states are not true four-body Efimov states since there is an $HL$ bound state,
but they can be regarded as three-body Efimov states of $H\!L$+$H$+$H$.
They are indicated
with the notation $(H\!L)H_2$ in Fig.~\ref{Fig:4BSpec} where the $H_2L$+$H$
and $H\!L$+$H$+$H$ thresholds intersect since $a_{H\!H}^*$$\rightarrow$$\infty$ at these points. 

Figure~\ref{Fig:4BSpec} sketches the energy trajectories for $H_3L$ as a function of $a_{H\!L}$.
In addition to the $(H\!L)H_2$ Efimov states, there are the $H_2L$ 
 Efimov states, and associated with
each of these we find --- for both $M/m$=50 and $M/m$=30 --- one $H_3L$
state that appears to be the analog of the universal four-boson states in 
Refs.~\cite{4BPlatter2004,4BHammer2007,4BStecher2009,4BFerlaino2009,Schmidt2010}. We  
find that its binding energy is universally related to the binding energy of the associated $H_2L$ Efimov 
state by $E_{H_3L}/E_{H_2L}$$\approx$$0.4$ for both of the mass ratios we have calculated. 
Note that the 
binding energies are defined relative to the next lowest breakup threshold ($H\!L$+$H$+$H$ for $H_2L$ states and $H_2L$+$H$ for $H_3L$ states).

\begin{figure}
\includegraphics[clip=true,width=\columnwidth]{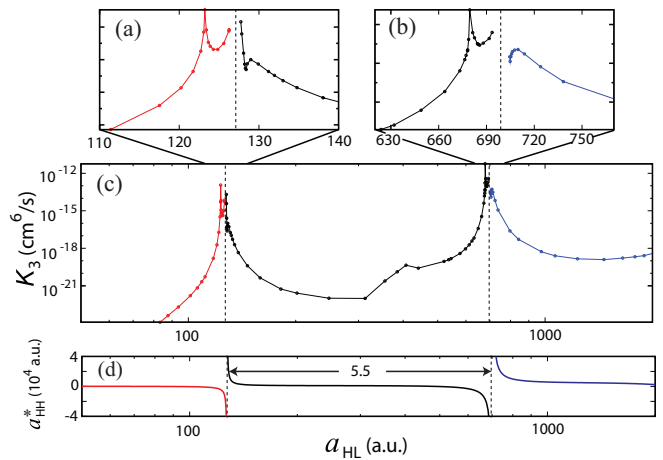}
\caption{
$K_3$ for $H\!L$+$H$+$H$$\rightarrow$$H_2L$+$H$ 
when $M/m$=30. 
The main peaks in (c) reflect the Efimov physics of $H_2L$ and occur where (d) $a_{H\!H}^*$ diverges.
The expanded plots (a) and (b) highlight the features on each main peak related to $H_3L$ states.
The vertical dashed lines indicate where $a_{H\!H}^*$=$\infty$.
}
\label{Fig_Recomb_30}
\end{figure}

Figure~\ref{Fig_Recomb_30} shows our numerically calculated 
three-body recombination rates $K_3$ for $H\!L$+$H$+$H$$\rightarrow$$H_2L$+$H$ with $M/m$=30
(see Ref.~\cite{HeCalc2} for details of our numerical methods).
Although there may be several final $H_2L$ 
Efimov states available, 
our calculation shows that recombination into 
the most weakly bound Efimov state dominates.
In fact, the main peaks in Fig.~\ref{Fig_Recomb_30}(c) occur where an $H_2L$
Efimov state just becomes bound.  
The separation between the main peaks is thus determined by 
$s_0$ from the $H_2L$ Efimov effect:
for $M/m$=30, $a_{H\!L}^{(2)}/a_{H\!L}^{(1)}$=$e^{\pi/s_0}$=4.34.
Note that the adiabatic hyperspherical approach gives $e^{\pi/s_0}$=3.96~\cite{Efimov1973,MassDIncao2006},
giving an indication of the BO approximation error for this mass ratio.  For consistency, we will quote only BO 
results in the rest of this Letter.
The factor of 5.5 between these two main peaks in Fig.~\ref{Fig_Recomb_30} does not match this 
prediction 
because the criterion 
$a_{H\!L}$$\gg$$ r_0$ is not well satisfied.

\begin{figure}
\includegraphics[clip=true,width=\columnwidth]{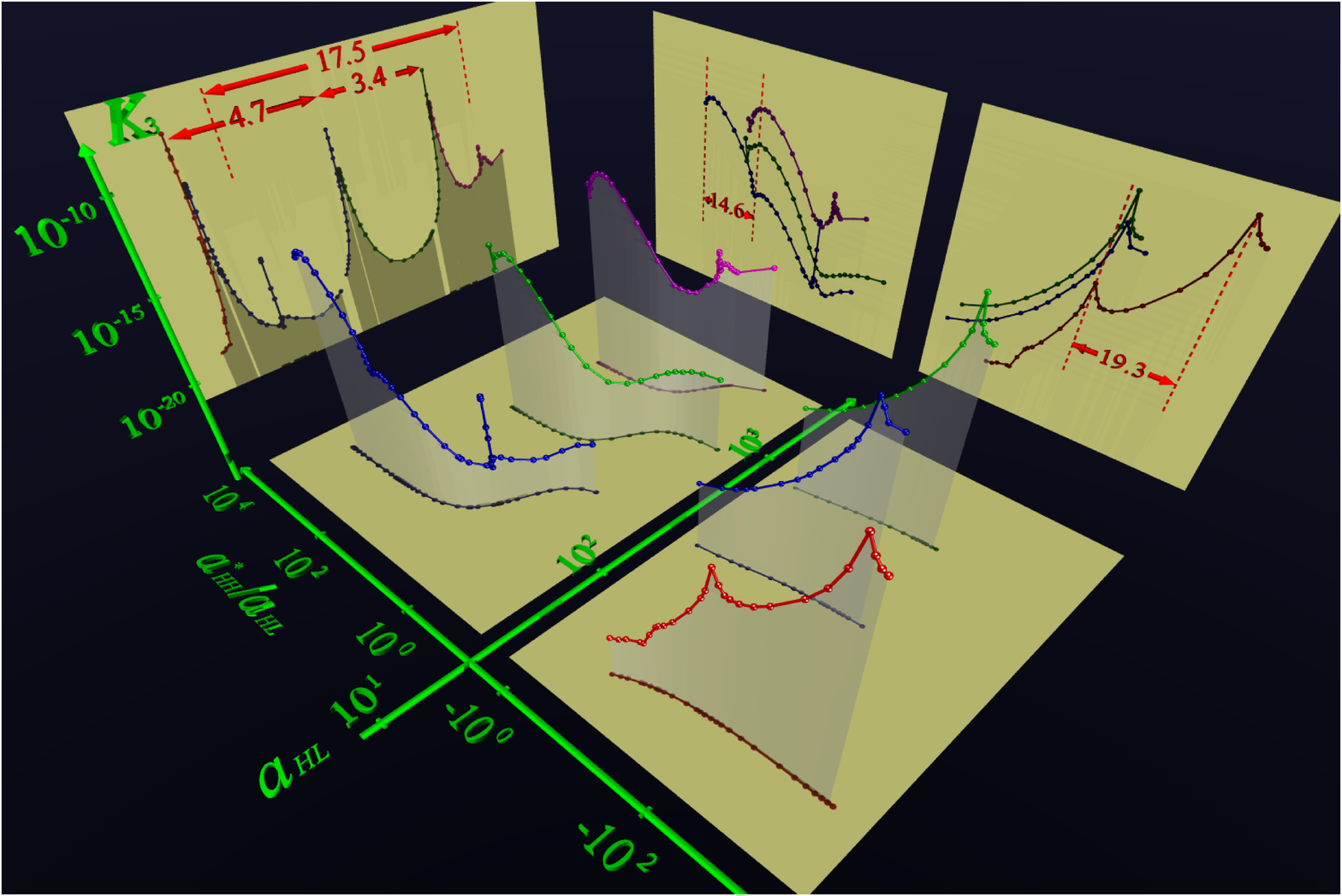}
\caption{
$K_3$ for $H\!L$+$H$+$H$$\rightarrow $$H_2L$+$H$ 
with $M/m$=$50$ simultaneously showing the
dependence on $a_{H\!L}$ and on $a_{H\!H}^*$.  The 
$H_2L$
Efimov physics is highlighted
by the $a_{H\!L}$-$K_3$ projection while the $a_{H\!H}^*$-$K_3$ projection illustrates the 
$(HL)H_2$
Efimov physics.
Note that $a_{H\!H}^*$ increases into the figure.
}
\label{Fig_Recomb_50}
\end{figure}

Since each main peak corresponds to a pole of $a_{H\!H}^*$, 
$K_3$ shows Efimov features characteristic of the $H_3$ motion where
$|a_{H\!H}^*|$$\gg$$a_{H\!L}$.  When this condition is satisfied,
the rates are given by the usual universal three-body expressions~\cite{Braaten2006,JoseScaling}
with the short-range length scale set to $a_{H\!L}$
\begin{align}
K_3^{(a_{H\!H}^*<0)}&=\frac{C'}{\mu}\frac{\sinh(2\eta)\,|a_{H\!H}^*|^4}{\sin^2[g_0\ln(\lvert a_{H\!H}^*\rvert/a_{H\!L})\!+\!\Phi']\!+\!\sinh^2(\eta)} \nonumber\\
K_3^{(a_{H\!H}^*>0)}&=\frac{C}{\mu}\sin^2[g_0 \ln(a_{H\!H}^*/a_{H\!L})+\Phi]\,({a_{H\!H}^*})^4.
\label{K3}
\end{align}
In these expressions, $C$ and $C'$ are universal constants. But, because the final $H_2L$
state is an
Efimov state, $\Phi$, $\Phi'$ and $\eta$ depend not on short-range four-body physics, but rather 
on the short-range
physics of the $H_2L$
states --- no additional four-body parameter is needed~\cite{4BPlatter2004,4BHammer2007,4BStecher2009}.  The $a_{H\!H}^*$-$K_3$ projection in 
Fig.~\ref{Fig_Recomb_50} supports this conclusion, showing that the  positions 
of the Efimov features related to the $H_3$ motion described by Eq.~(\ref{K3}) are approaching a
universal position in the limit 
$a_{H\!L}$$\gg$$r_0$.

Interestingly, for larger mass ratios, $H_2L$ Efimov states with non-zero orbital angular momentum $j$ 
are possible~\cite{Efimov1973,MassDIncao2006}. 
In this case, the universal constant $s_0$ in Eq.~(\ref{Eq_EfimovScale}) is determined
from $s_0^2$=$\chi_0^2M/2m\!-\!j(j\!+\!1)\!-\!1/4$, and there will be an Efimov effect for $H_2L$ so long as $s_0^2$$>$0. 
Higher angular momentum Efimov states have not yet been observed because experiments 
have focused mostly on identical particles which have no such states
and because $j$$>$0 Efimov states produce extremely narrow recombination peaks as seen in Fig.~\ref{Fig_Recomb_50}.
The mass ratio in Fig.~\ref{Fig_Recomb_50}, $M/m$=50, 
 supports $j$=2 Efimov states, 
but is not sufficient for higher $j$ states.
Because $l_{\rm eff}$=$j$$\neq$0 for the $H_2L$ system, there is no $(HL)H_2$ Efimov effect associated with
these states.  
Their narrow $K_3$ peaks thus show no substructure of the
sort seen on the main peaks.  

The periods for each of the three different families of Efimov peaks are indicated in Fig.~\ref{Fig_Recomb_50}.
Because none of the respective scattering lengths are strongly in the universal limit, however,
the calculated periods do not match the predicted ones. 
The main $j$=0 $H_2L$
Efimov peaks highlighted by the $a_{H\!L}$-$K_3$ projection should have a period of $e^{\pi/s_0}$=3.08.  The calculated spacings
are larger than this, but appear to be approaching the expected value as $a_{H\!L}$
increases.  Similarly, for $j$=2, the expected period is 10.46 while the calculated one is 17.5, and
the predicted period of the $(HL)H_2$ Efimov substructure on the main peaks 
highlighted in the $a_{H\!H}^*$-$K_3$ projection is $e^{\pi/g_0}$=22.7.
The magnitude of the deviations from the predictions likely reflects what can be observed since it is difficult to
penetrate deeply into the universal regime experimentally.

So far, we have not taken advantage of all of the freedom that this heteronuclear
system affords to manipulate the Efimov features.
In particular, since $a_{H\!H}^*$ depends only on the total $H$+$H$ interaction, i.e. effective plus direct interactions, 
it can be tuned by either interaction --- or both.  Intriguingly, this freedom also allows 
experimental control of both $a_{H\!H}^*$ and $\Phi$ (or $\Phi'$).  Controlling the former
via either $a_{H\!L}$ or a direct interaction with scattering length $a_{H\!H}$ allows
the various Efimov features seen, for instance, in Figs.~\ref{Fig_Recomb_30} and \ref{Fig_Recomb_50} to be mapped out.
Controlling the latter makes it possible to shift all of these features --- something not
possible so far in three-body systems.  Moreover, in the neighborhood of a pole in
$a_{H\!H}^*$, it should be possible to exert both types of control largely independently.

\begin{figure}
\includegraphics[clip=true,width=0.95\columnwidth]{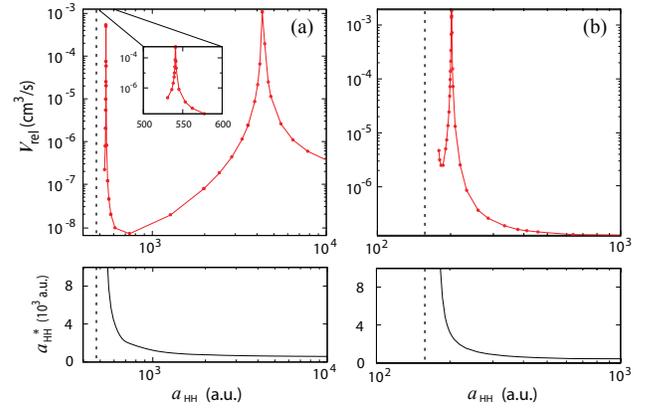}
\caption{
$V_{\mathrm{rel}}$ for $H_2L(n)$+$H$$\rightarrow$$H_2L(n-1)$+$H$ for  (a) $n$=1
 and (b) $n$=2. 
The dashed lines indicate the positions where $|a_{HH}^*|$=$\infty$.
}\label{Fig_Relax_30}
\end{figure}

To illustrate the effect of tuning the direct interaction, we show in Fig.~\ref{Fig_Relax_30} the numerically
calculated rates for the relaxation process 
$H_2L(n)$+$H$$\rightarrow$$H_2L(n-1)$+$H$ as a function of $a_{H\!H}$ where $n$ labels the most weakly
bound $H_2L$ state.  This tuning was accomplished by including a short-range, direct $H$+$H$ interaction
in addition to an effective $H$+$H$ interaction.
The behavior of $V_{\rm rel}$ in Fig.~\ref{Fig_Relax_30} is also found in the three-boson system.  In fact, when $a_{HH}^*\gg a_{HL}>0$, $V_{\mathrm{rel}}$
has the same form as in three-boson systems
~\cite{Braaten2006}
\begin{eqnarray}
V^{(a_{H\!H}^*>0)}_{\rm rel}\!\!\!&=\!\!\!&\frac{A \sinh 2\eta}{\sin^2[g_0\!\ln(a_{H\!H}^*/a_{H\!L})\!+\!\Phi]\!+\!\sinh^2\!\eta}a_{H\!H}^*,
\label{Eq_V}
\end{eqnarray}
where $A$ is a universal constant, but $\eta$ and $\Phi$ depend on the short-range details of the relaxed $H_2L$ bound state ---
although not on a separate four-body parameter.

To conclude, we have studied Efimov physics in the four-body heteronuclear system $H_3L$
with bosonic $H$ atoms, showing that there is no true four-body Efimov effect.  We have, however, identified a universal four-body state linked to each $H_2L$ Efimov state.
Perhaps more significantly, we have shown that the richness of the heteronuclear system can
be exploited to reveal a variety of Efimov features with different universal scaling by scanning just $a_{H\!L}$.
That same richness allows more opportunities for controlling the system as well, which is potentially
very important for experiments seeking to study or utilize these processes. For instance, 
taking advantage of the BO approach introduced in Ref.~\cite{Fonseca1981}, we pointed out that scanning
$a_{H\!L}$ gives a means for controlling the scattering length between the heavy atoms --- a
tool that could prove useful experimentally.  
Heteronuclear four-body and larger systems thus provide considerable possibilities for 
studying Efimov physics that were not previously anticipated.

\begin{acknowledgments}
We thank C.H. Greene for critical and stimulating discussions of the four-body Efimov effect. 
This work was supported in part by the National Science Foundation and in part by the
Air Force Office of Scientific Research.
Y. W. also acknowledges support from the National Science Foundation under Grant No. PHY0970114.
\end{acknowledgments}

\end{document}